\documentclass[submission,copyright,creativecommons]{eptcs}
\providecommand{\event}{FESCA 2014}

\usepackage[usenames,dvipsnames]{color}
\usepackage[T1]{fontenc}
\usepackage{stmaryrd}
\usepackage{pifont}
\usepackage{alltt}
\usepackage{amssymb}
\usepackage{amsmath}
\usepackage{float}
\usepackage{multirow}
\usepackage[pdftex]{graphicx}
\usepackage{verbatim}
\usepackage{colortbl}
\usepackage{xspace}
\usepackage{tikz}
\usepackage{xypic,url,subfig,pdflscape}

\newcommand{\somespace}{\quad\quad}

\newcommand{\ie}{\textsl{i.e.,}\xspace}
\newcommand{\eg}{\textsl{e.g.,}\xspace}

\floatstyle{ruled}
\newfloat{program}{thp}{lop}[section]
\floatname{program}{Program}

\def\titlerunning{Comprehensive Monitor-Oriented Compensation Programming}
\def\authorrunning{C. Colombo and G. J. Pace}

\title{Comprehensive Monitor-Oriented Compensation Programming}
\author{Christian Colombo\qquad Gordon J. Pace
\institute{Department of Computer Science\\ University of Malta}
\email {\{christian.colombo $|$ gordon.pace\}@um.edu.mt}  }

\begin{document}

\maketitle

\begin{abstract}
Compensation programming is typically used in the programming of web service compositions whose correct implementation is crucial due to their handling of security-critical activities such as financial transactions. While traditional exception handling depends on the state of the system at the moment of failure, compensation programming is significantly more challenging and dynamic because it is dependent on the runtime execution flow --- with the history of behaviour of the system at the moment of failure affecting how to apply compensation. To address this dynamic element, we propose the use of runtime monitors to facilitate compensation programming, with monitors enabling the modeller to be able to implicitly reason in terms of the runtime control flow, thus separating the concerns of system building and compensation modelling. Our approach is instantiated into an architecture and shown to be applicable to a case study.
\end{abstract}

\section{Introduction}

With the advent of long-lived transactions, particularly in the context of web service compositions, compensation programming has gained considerable attention from both industry and academia. While the \emph{de facto} standard of programming business processes having compensations is BPEL \cite{bpel2}, the literature (see \cite{survey} for an overview) is inundated with varying proposals of formal approaches to compensation programming. 
The main challenge which these attempt to address is the dynamic aspect of compensations; compensations are chosen depending on the control-path being taken, i.e.\ a snapshot of the history of the system at the moment of failure, which is unknown at compile time. For example the cancellation of an online transaction depends on whether the payment has \emph{previously} been affected or not, whether the transport arrangement has \emph{earlier} been confirmed or not, etc. This is significantly different from traditional exception handling in the \emph{try-catch} fashion since the latter is generally static, i.e.\ depends on the state of the system at the moment of failure, e.g., on an invalid user input, or corrupt data being read from a file. 

Yet, even vanilla exception handling may need to reason about history snapshots --- something which typical programming languages do not support well. For this reason, software monitoring techniques such as monitor-oriented programming (MOP) \cite{MJG+11mop}, have been specifically proposed to support reasoning about control paths by detecting events such as method calls and matching event patterns to logic formulae. 

Note that while compensations historically have roots in error recovery in highly concurrent contexts \cite{DAVI72,RICSD78}, nowadays compensations are also commonly used to model exceptional case handling as part of the normal flow of a system. This is also the case with MOP; it originates from a background of error recovery techniques but can be used to trigger any functionality through monitoring. Similarly, as the examples we use in this paper, we do not confine ourselves to the error handling aspect of compensation programming and indeed our approach can be used alongside exception handling techniques.

When MOP is used to support error handling, it can be used to help the programmer detect exceptions, but it does little to support their reparation. Put differently, MOP is able to detect \emph{when} an exception has occurred but does not directly support \emph{how} to handle the exception. 
This is not surprising in the context of exception handling because the programmer is typically given complete freedom as to the choice of the recovery code. However, this is not the case with compensation programming \cite{survey} because it exploits the execution pattern of a program to deduce the appropriate recovery action:  
$(i)$ knowledge of what has been executed enables the demarcation of actions which have to be compensated for; while 
$(ii)$ knowledge of the execution pattern enables the composition of the corresponding compensation actions (typically in reverse order of their original execution). 
This has the advantage of partially automating the \emph{how} of compensation programming while giving rise to two further questions: \emph{what} to compensate for, and \emph{which} compensation strategy to use if there is more than one way of compensating an earlier behaviour. 
More concretely, the four questions of \emph{when}, \emph{what}, \emph{how} and \emph{which} can be understood in terms of the following example: 

Consider an e-procurement system through which users are able to login, order, and pay for bought items which are subsequently shipped to the customer. During the transaction, the user can also cancel the order. A possible trace of actions is:  


\medskip\medskip
\begin{center}
\scalebox{0.8}{\includegraphics{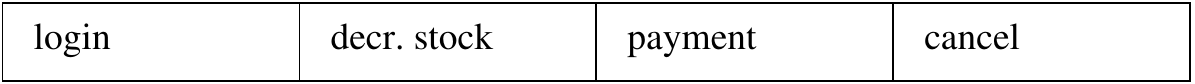}}
\end{center}
\medskip

\begin{description}
\item[\emph{When} to trigger compensations.]
At each step of a process, compensations can potentially be triggered. In the case of the example trace, compensations may be triggered \emph{when} the user cancels the purchase (shown below by \emph{comp} in the top right corner). 
Note that while compensations are typically triggered upon an exception throw (e.g., payment fails) as an error handling mechanism, they may also be used for programming normal business logic as in the case of this example. 

\medskip\medskip
\begin{center}
\scalebox{0.8}{\includegraphics{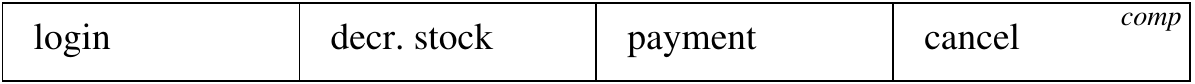}}
\end{center}
\medskip

\item[\emph{What} to compensate for.] 
When compensating, due to the correspondence between actions and their compensations, one would have to deduce \emph{what} actions to compensate for. Compensations by definition compensate for previously completed actions (e.g., a completed payment). However, not all previous actions might need to be compensated for. For example the login action (see figure below) does not need to be compensated for when cancelling a purchase (the user would typically still want to remain logged in even though a purchase has been cancelled). 

\medskip\medskip
\begin{center}
\scalebox{0.8}{\includegraphics{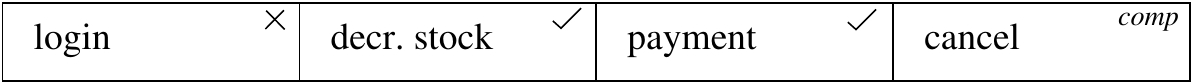}}
\end{center}
\medskip

\item[\emph{How} to compensate for actions.]
Building upon the example, for each activity earmarked for compensation, an action has to be selected for the job. As shown in the figure below there may be one or more ways in which an activity may be compensated for. To compensate for a decrement in stock, one would simply increment the stock while a payment is compensated in full or in part depending on the context. 

\medskip\medskip
\begin{center}
\scalebox{0.8}{\includegraphics{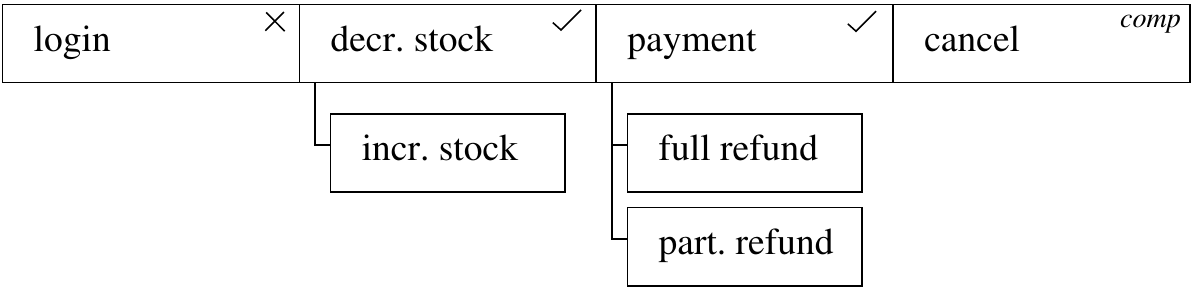}}
\end{center}
\medskip

\item[\emph{Which} compensation strategy to use.]
The next step is to choose \emph{which} of the possible compensation options are to be executed. In the example (depicted below) the choice of the type of refund depends on whether the user is a privileged user or a standard user at the time of refund (which might have changed since the time of payment). In this case, we have opted for the partial refund, assuming the user is standard. Once the compensation actions are outlined, these are typically executed in the reverse order of the corresponding actions. 
A similar issue is that of the order in which compensations are to be performed. To be true to the notion of compensations, sequential behaviour is typically compensated for in reverse order, but concurrent behaviour may be compensatable in a concurrent manner.
These decisions merit abstraction and careful consideration when programming compensations. 

\medskip\medskip
\begin{center}
\scalebox{0.8}{\includegraphics{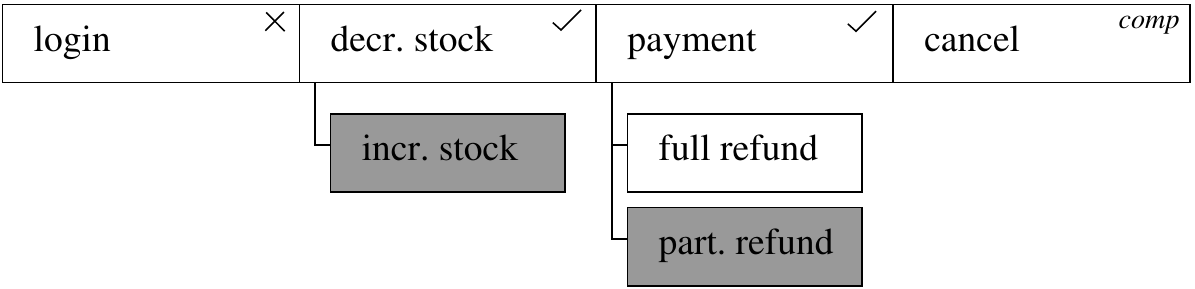}}
\end{center}
\medskip

\end{description}

The simple example above highlights the number of questions involved in compensation programming. While we have only considered the choices on a per-action basis, in real life scenarios the decisions might need to be taken on sub-sequences of actions. For example a sub-sequence of actions such as the four actions in the above scenario might have a specific single compensation, logout, if the user is detected to be fraudulent. Furthermore, coming up with answers to the questions may involve arbitrary complex logic considering various issues including the control flow, system performance, fraud, and any relevant attributes of the entities involved in the transaction. These intricacies motivate the need to abstract these questions and use an appropriate design to handle their interaction. 

Given the use of runtime monitors for the detection of complex patterns of system behaviours as in MOP, applying monitoring to these questions seems to be a natural solution. 
In previous work \cite{cp13cas}, we have proposed a monitoring approach, monitor-oriented compensation programming (MOCP) to the programming of the \emph{what} and \emph{how} questions through a compensation manager residing on a monitoring component receiving live updates from the system. 
Our contribution in this paper is to extend the architecture so as to enable the programming of all four issues in compensation programming. This is done by combining standard runtime monitoring to seamlessly interact with the compensation manager, instructing it when and which compensations to trigger --- signals, which in previous work, had to be provided by the system itself. Such an arrangement enables the specification of compensations to be done at a higher level of abstraction --- in the monitoring language --- while also providing more fine-grained separation of concerns, alleviating the system of compensation programming. 

In the rest of the paper, we start by providing the preliminaries in Section \ref{sec:prelim} and go on to introduce the proposed architecture in Section \ref{sec:arch}. Subsequently, we show how our approach is applicable to a case study in Section \ref{sec:cs} followed by a more practical account of how the architecture can be applied to real-life scenarios in Section \ref{sec:prac}. Finally, we compare our proposal to related work in Section \ref{sec:rw}, whilst concluding in the final section.

\section{Preliminaries}\label{sec:prelim}

\begin{figure}
\centering{\scalebox{0.8}{\includegraphics{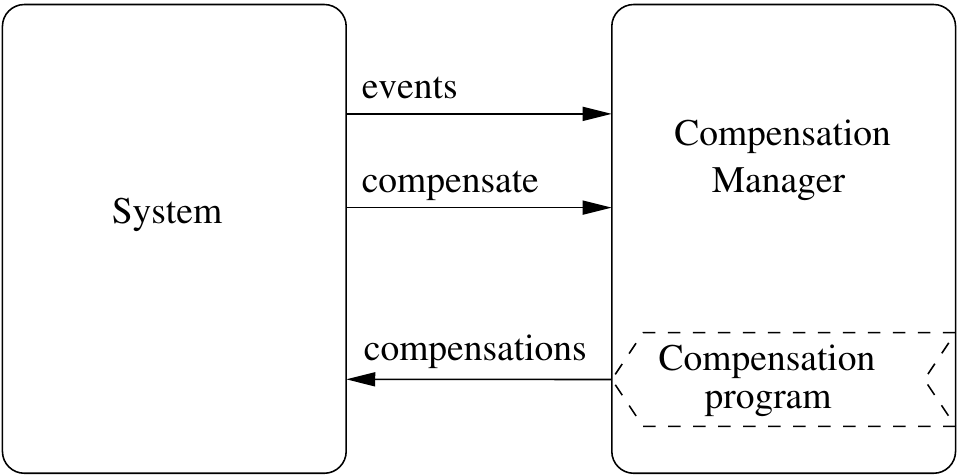}}}
\caption{The architecture with a compensation-aware system}
\label{fig:arch}
\end{figure}

In previous work \cite{cp13cas} we have proposed compensating automata --- a notation for specifying compensation logic. Through a case study, compensating automata have been shown to be able to express complex compensation logic alleviating the system from handling compensations. Instead, the system simply indicates the need to be compensated and the compensation manager triggers compensations. Figure~\ref{fig:arch} depicts monitor-oriented compensation programming as previously proposed, where the system communicates all the relevant events to the compensation manager while the latter decides \emph{what} events to compensate for and \emph{how} --- collating the applicable compensations. If the system reaches a state where it needs to execute compensations, the system signals the compensation manager which in turns starts executing the collated compensations. Upon executing all the relevant compensations, the compensation manager returns control back to the system.  

\begin{figure}
\centering{\scalebox{0.8}{\includegraphics{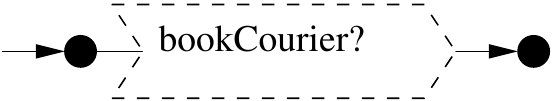}}}\\
\medskip\medskip\medskip
\centering{\scalebox{0.8}{\includegraphics{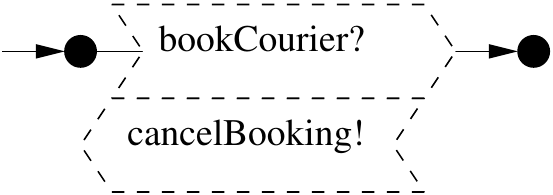}}}\ \ \ \ \ \ \ \ 
\centering{\scalebox{0.8}{\includegraphics{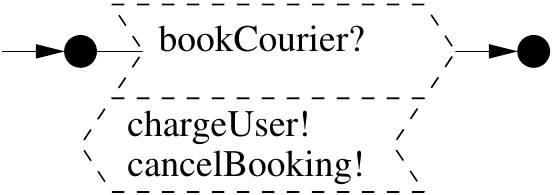}}}\\
\caption{Basics of compensating automata}
\label{fig:ca-basic}
\end{figure}

To assist the user in programming the compensation manager, we have proposed compensating automata --- essentially state machines which listen to system events and compose compensations accordingly. A basic transition of a compensating automaton listens for an event and installs no compensation (Figure~\ref{fig:ca-basic} [top]). To program a compensation installation upon an event, one or more compensation instructions can be specified as shown in Figure~\ref{fig:ca-basic} [bottom].

\begin{figure}
\centering{\scalebox{0.8}{\includegraphics{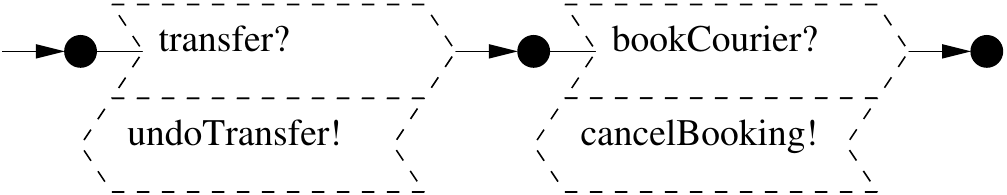}}}\\
\medskip\medskip\medskip
\centering{\scalebox{0.8}{\includegraphics{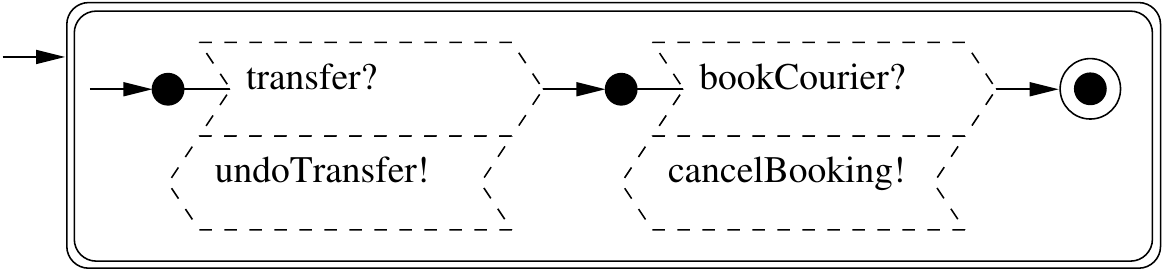}}}\\
\caption{Sequencing and scoping in compensating automata}
\label{fig:ca-advanced}
\end{figure}

Furthermore, transitions can be composed in sequence as shown in Figure~\ref{fig:ca-advanced} [top], signifying that subsequent compensations are installed on a stack and executed in reverse order if activated. In particular cases, the user might need to purge the stack and this can be programmed in terms of the box (shown in Figure~\ref{fig:ca-advanced} [bottom]) such that when execution reaches the end of the box, the stack with the compensations collected during the box's execution is purged and discarded. Note that the box is double edged in this case because it is also acting as the final (and only) state of the parent automaton. 

While compensating automata provide further advanced features for compensation programming, this short introduction suffices to help the reader understand the case study presented in Section~\ref{sec:cs}. The next section introduces the modified MOCP architecture, extending the role of monitoring further from our previous work \cite{cp13cas}. Note that in the rest of the paper MOCP will refer to the extended architecture.

\section{A Compensation-Handling Architecture}\label{sec:arch}

To support comprehensive compensation programming --- alleviating a system from compensation programming up to the point where it need not be aware of compensations --- the proposed architecture is entirely based on monitoring. 
We decompose this architecture into two parts: 
$(i)$~one which deals with compensation aspects such as compensation installation and discarding, addressing the questions of \emph{what} and \emph{how} to compensate, and 
$(ii)$~another which deals with triggering compensations at particular points in time, addressing \emph{when} and \emph{which} compensation strategy to execute. Note that the former is involved directly with compensation programming while the latter simply triggers compensation strategies of choice at particular points in time. 
The first component of such an architecture can be programmed using compensating automata which have been specifically devised for such programming. On the other hand, encoding the \emph{when} and \emph{which} logic of compensations can be done by any specification notation which can be used for runtime verification and supports the triggering of reparatory actions. For this job we choose Dynamic Automata with Timers and Events (DATEs) \cite{CGG08FMICS} because they are somewhat natural to integrate with compensating automata --- they are also automaton-based and support channel communication. Note that the proposed architecture can also be instantiated using any other suitable specification notation. 

\begin{figure}[t]
\centering{\scalebox{0.8}{\includegraphics{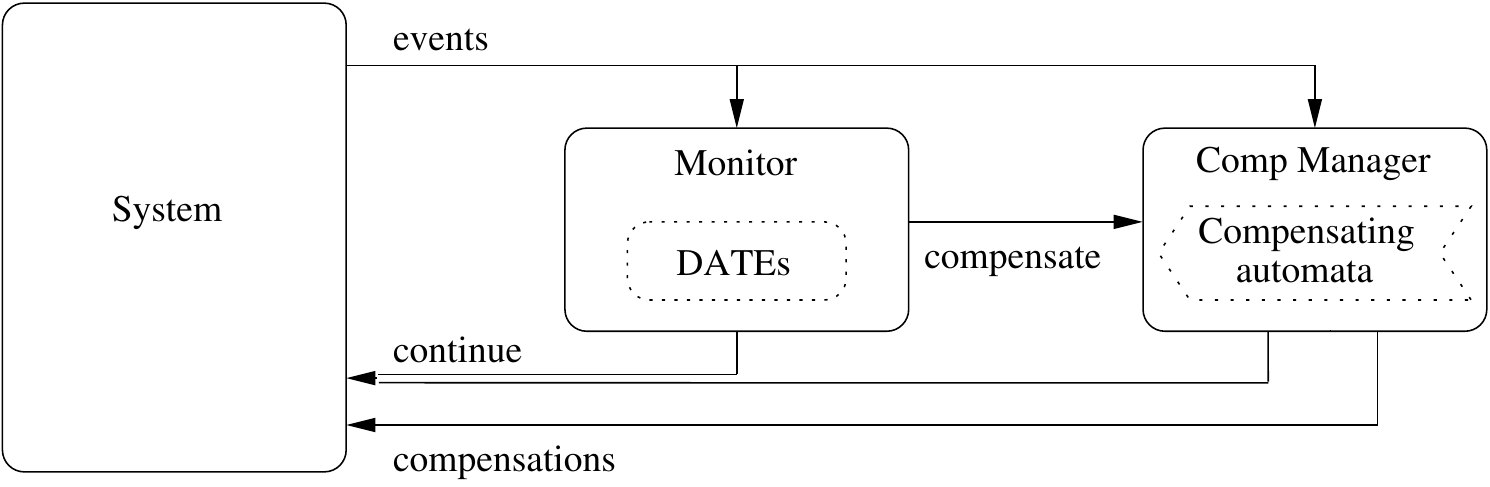}}}
\caption{The monitor-oriented compensation programming architecture}
\label{fig:mocp}
\end{figure}

In the proposed architecture, as shown in Figure~\ref{fig:mocp}, the monitoring components may interact with the system through three connections: one which enables the system to communicate events, another enabling the monitor to signal the system to continue, and a third one on which the compensation manager component instructs the system regarding what compensations are to be executed. Furthermore, the architecture is event-driven and each event the system emits triggers a series of steps as follows: 

\begin{enumerate}
\item The system emits an event and waits for the \emph{continue} signal from the compensation framework. 

\item The monitor receives the system event and processes it, emitting a \emph{compensate} signal to a particular compensation manager [answering the questions of \emph{when} and \emph{which} compensation to trigger] or a \emph{continue} signal (but not both).

\item The role of the compensation manager is the following:
\begin{enumerate}
\item It receives the system event and processes it, deciding whether to compensate for the event or not [answering the question of \emph{what} to compensate for] and in case of compensation, storing an appropriate compensating action [deciding the \emph{how} of compensations]. 
\item It then checks whether a \emph{compensate} signal has been received from the monitor.  If a \emph{compensate} signal has been received: 
\begin{enumerate}
\item\label{itm:emit} The compensation manager starts emitting compensations to the system. 
\item While the system is executing compensation instructions, the monitor and compensation manager are still potentially receiving and processing system events. 
\item When the compensation manager has sent all compensation instructions to the system (the stack is exhausted), it checks once more for the \emph{compensate} signal and if present repeats from Point~\ref{itm:emit}. Otherwise, execution continues as below. 
\end{enumerate}
\item The compensation manager issues a \emph{continue} signal. 
\end{enumerate}

\item Once the system detects two \emph{continue} signals 
it continues normally. 
\end{enumerate}

While the architecture seems to show a single \emph{compensate} line (Figure~\ref{fig:mocp}), in fact this line is multiplexed to enable the monitor to choose whichever compensation strategy is applicable. 
Similarly, although we show a single DATEs box, in effect there would typically be several DATEs and several compensating automata. 
Finally, we note that although the architecture is not formally defined beyond the algorithm given above, the components of the architecture, i.e.\ compensating automata and DATEs, have been fully formalised in previous work. 

The following section aims to showcase the benefits of this highly modular architecture through a case study.  More concrete details about the architecture are subsequently provided in Section~\ref{sec:prac}.


\newcommand{\cont}{(cont.)}

\section{An eProcurement Case Study}\label{sec:cs}

The use of MOCP as proposed in this paper is particularly useful when the decision to activate a compensation strategy is not straightforward. One example where this is the case is an implementation of an e-procurement system\footnote{Inspired from the e-procurement system presented in \cite{GFJK03notenough} and the Entropay system presented in \cite{CPA12async}.} 
which handles payments and shipments of goods. The e-procurement system allows users to create virtual credit cards and subsequently (the emphasised words are system events observable by the monitors): 
\begin{itemize}
\item
\emph{Load} money onto the virtual credit cards from their personal bank accounts. 
\item
\emph{Transfer} money across virtual credit cards. 
\item
\emph{Order} goods from a third party using the e-procurement system as an intermediary. This involves concurrently \emph{paying} the third party through the bank and \emph{booking a courier} (either courier \emph{A} or \emph{B} depending on availability).
\end{itemize}

In this case study, we assume that the system as described above is already implemented and that the relevant events can be detected by a monitor when they occur. Note that there is no mention of compensations in the above description since all compensation logic will be handled by the monitor in a MOCP fashion. 
In the rest of this section we thus elaborate on the monitor logic which handles compensation programming and triggering. 
Finally, we conclude the section with a discussion of the case study.

\subsection{Compensation Logic for the eProcurement Case Study}

During the normal activities of the eProcurement system described above, a number of possible failures might occur (elaborated further below) causing the transaction to fail. Cancelling a courier or reversing bank transactions usually incur a charge and under certain circumstances it might not be possible to cancel such operations at all, \eg when the shipment has already left, (corresponding to \emph{what} to compensate for). Furthermore, assuming that compensation is possible, there are at least three parties who might incur the charge (corresponding to the \emph{how} of compensations): the user, the third party involved (\eg the courier or the bank), or the intermediary, i.e.\ the e-procurement system. This means that for one courier booking transaction, there are at least three possible compensation strategies (corresponding to \emph{C1}, \emph{C2}, and \emph{C3} in Figure~\ref{fig:cs-cas}) whose choice can only take place after the error actually occurs (as it depends on the kind of error). 
Note that in all cases compensations are discarded with no replacement once the shipment of goods occurs, \ie the booking cannot be cancelled after shipment. 

Similarly, for the banking transactions there are at least three possible compensations plus a fourth one which keeps track of the credit cards used so that these can be blocked in case of blacklisted users (corresponding to \emph{B1}, \emph{B2}, \emph{B3}, and \emph{B4} in Figure~\ref{fig:cs-cas}). Note that in this case, once payment succeeds, we assume that the money loads and transfers are not to be undone. 

\begin{figure}
\centering
\subfloat[Monitoring for bank errors]{\label{sfig:mon1}
\centering{\scalebox{0.8}{\includegraphics{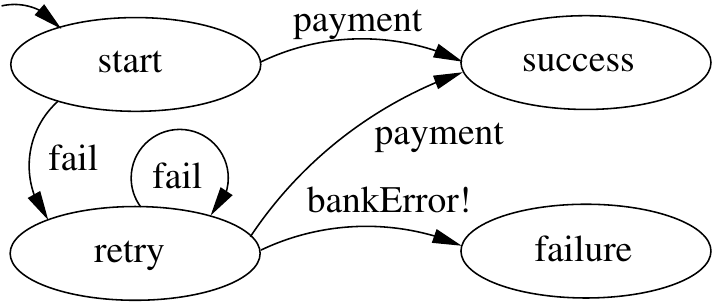}}}
}\ 
\subfloat[Monitoring for courier errors]{\label{sfig:mon2}
\centering{\scalebox{0.8}{\includegraphics{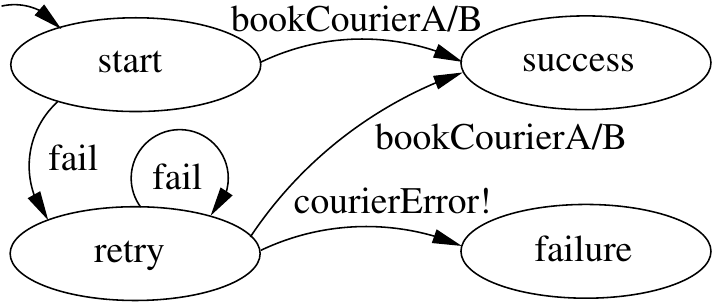}}}
}\\
\medskip\medskip
\subfloat[Monitor for whitelisting and blacklisting users]{\label{sfig:mon3}
\centering{\scalebox{0.8}{\includegraphics{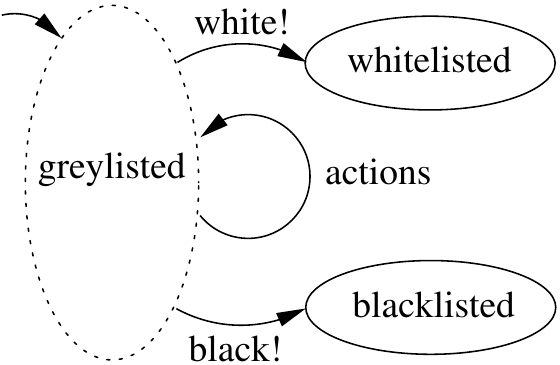}}}
}\\
\medskip\medskip
\subfloat[Receiving signals for other monitors and triggering compensation strategies]{\label{sfig:mon4}
\centering{\scalebox{0.8}{\includegraphics{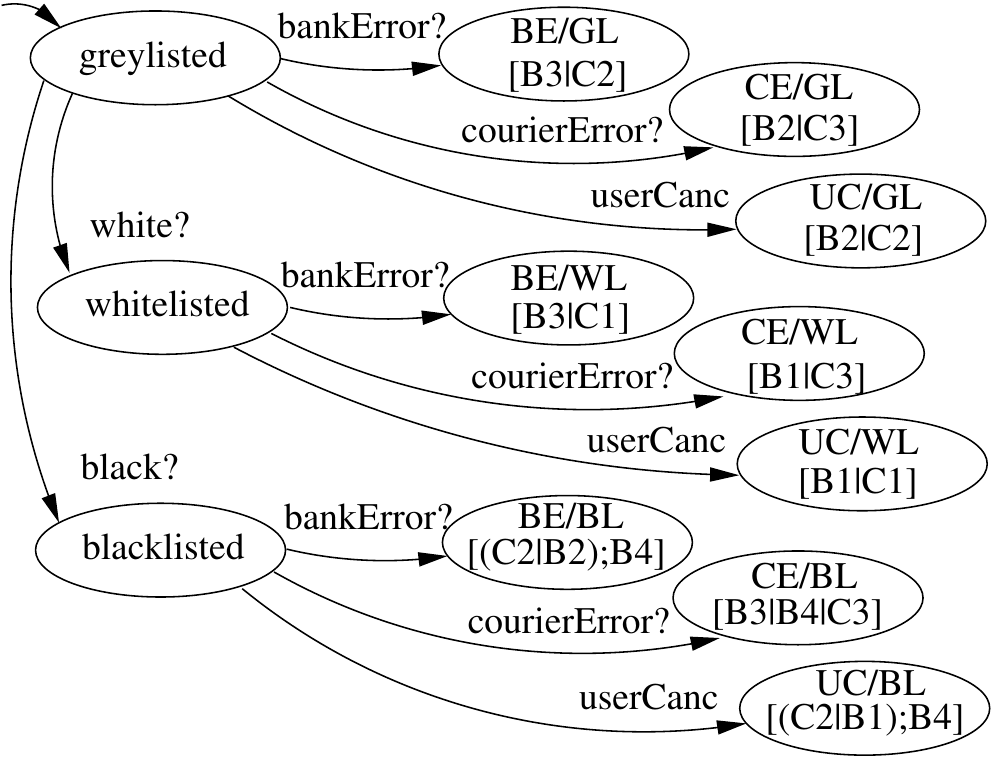}}}
}\\
\caption{Monitoring for triggering compensations for an e-procurement system}
\label{fig:cs-mons}
\end{figure}

To decide \emph{which} compensation strategy to trigger \emph{when}, the e-procurement system classifies users and errors so that the charge is incurred by different parties under different circumstances: 

{\bf Classification of errors}
The e-procurement system policy broadly considers three kinds of errors: a bank error, a courier error, or a user cancellation. In the case of a user cancellation it is always the user who should pay for cancellation charges. However, in case of a banking error, the cancellation charges for the banking transactions are incurred by the bank while the cancellation charges for the courier transaction are incurred by the user or the e-procurement system depending on the type of user (see next point). A similar approach is taken in case of a courier error. Failure detection is handled by the two (simplified) monitors depicted in Figure~\ref{fig:cs-mons}\subref{sfig:mon1} and \subref{sfig:mon2}. These monitors listen for system events such as \emph{payment} or \emph{bookCourier} (more insight into event capturing is given in the next section). If after a number of retries the system fails to perform the successful event, the monitor communicates \emph{bankError} or \emph{courierError} to the main monitor shown in \subref{sfig:mon4}.

{\bf Classification of users}
Users are classified as whitelisted, greylisted, or blacklisted. Blacklisted users are suspicious users who cannot be trusted. For this reason these users are not automatically given back their money in case of a failure. Instead, their credit cards are blocked till after human investigation. 
On the other hand, whitelisted users are trusted customers for whom certain allowances are made such as paying cancellation charges on their behalf. 
Greylisted users are users who are neither blacklisted nor whitelisted. Each user starts off as greylisted and at a particular point in time a designated monitor (partly shown in Figure~\ref{fig:cs-mons}\subref{sfig:mon3}) might classify the user as blacklisted or whitelisted.


Note that such an e-procurement strategy for handling cancellation includes nine different compensation strategies depending on three user types and three kinds of failure. Clearly, deciding the compensation strategy for such a scenario is not straightforward. By separating the different aspects of compensation programming, the decision can be taken using monitors as depicted in Figure~\ref{fig:cs-mons}\subref{sfig:mon4} where compensation strategies (shown in square brackets) can be composed in parallel or sequentially depending on the case. For example, in case a blacklisted user cancels the transaction, the payment of the relevant charges should strictly occur before the credit cards are blocked. 
\vspace{-.1cm}

\subsection{Discussion}

The challenge of expressing non-trivial compensation logic in a modular and understandable fashion has been particularly highlighted by Greenfield et al.\ \cite{GFJK03notenough}, arguing that ``compensation is not enough''. 
As a solution the authors \cite{NFG+05sowl} propose an approach which is the complete
opposite of ours in terms of a model which does not differentiate between
normal and exceptional behaviour, and abstracts away from the notion
of compensations. They claim that this approach, which is based on a guarded-command
language, simplifies the specification of the system. Admittedly, the examples given in \cite{NFG+05sowl} are very readable since the model is somewhat similar
to a textual specification with many statements of the form: ``If this happens,
then this should happen''. The disadvantage of this approach is that the ``compensation
view'' is lost, i.e.\ the relationship between an action and its compensation,
the ordering of compensation execution, etc., is not visible from the
model. If such a view is not necessary for the programmer, then, indeed, this
model should be preferred. 
However, the compensation view has its advantages:
$(i)$ the correspondence between actions and their compensations is useful since
one would usually require part of the system state at the time of executing the
action to be available during the execution of the compensation; 
$(ii)$ compensation
notations usually clearly encode sequential constraints amongst actions and
compensations, information which is useful for ensuring temporal properties.
If these advantages are important in a given context, then employing compensating
automata might provide the right balance between the compensation view
and flexibility.

Using MOCP we have successfully shown how non-trivial compensation logic, comparable to that presented in \cite{GFJK03notenough}, can be programmed transparently to the underlying system in a modular fashion.

\section{Using the Architecture in Practice}\label{sec:prac}

While the previous section describes an account of how the architecture is applicable to a case study, in this section we elaborate on the practical steps which one would have to take to adopt the approach. Although the architecture has not been fully implemented, the monitor component has been fully implemented and has been previously applied independently to real-life financial transaction systems \cite{CGG08FMICS,CPA12async} written in Java, while the compensation manager is currently being implemented. In what follows we take the current implementation into consideration and comment on how it can be extended to make it fully functional.

\begin{figure}[t]
\centering{\scalebox{0.8}{\includegraphics{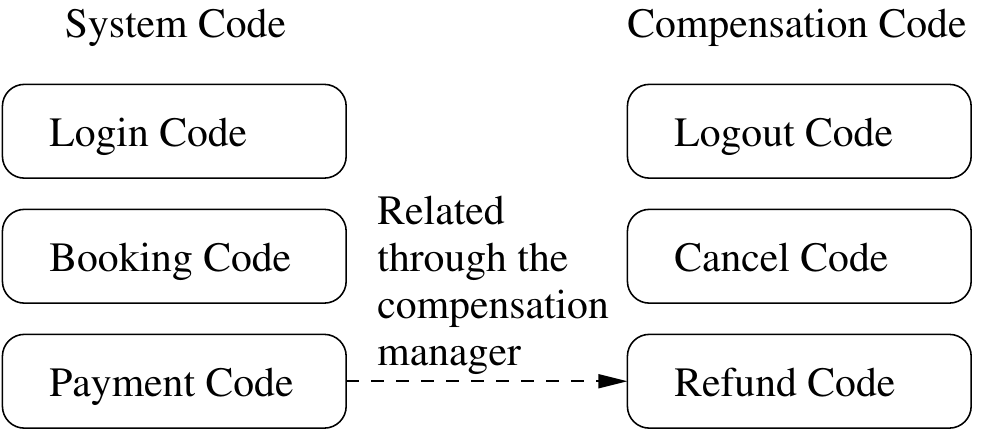}}}
\caption{The relation between system code and compensation code through the compensation manager}
\label{fig:prac}
\end{figure}

{\bf Programming the compensation code}
Due to the nature of compensations, which are not necessarily the exact reversal of the action being compensated for, the compensation code has to be coded by a programmer. However, the advantage of the proposed architecture is that the compensation code can be programmed in separate modules (see Figure~\ref{fig:prac}) and eventually executed by the compensation manager.

\begin{landscape}
\begin{figure}[t]
\centering{\scalebox{0.8}{\includegraphics{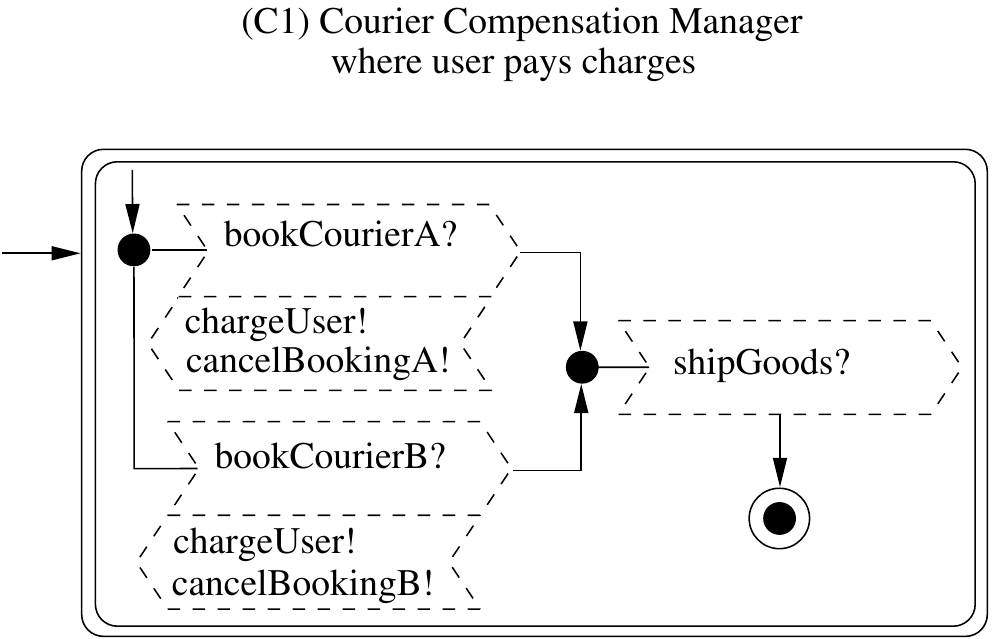}}}\somespace
\centering{\scalebox{0.8}{\includegraphics{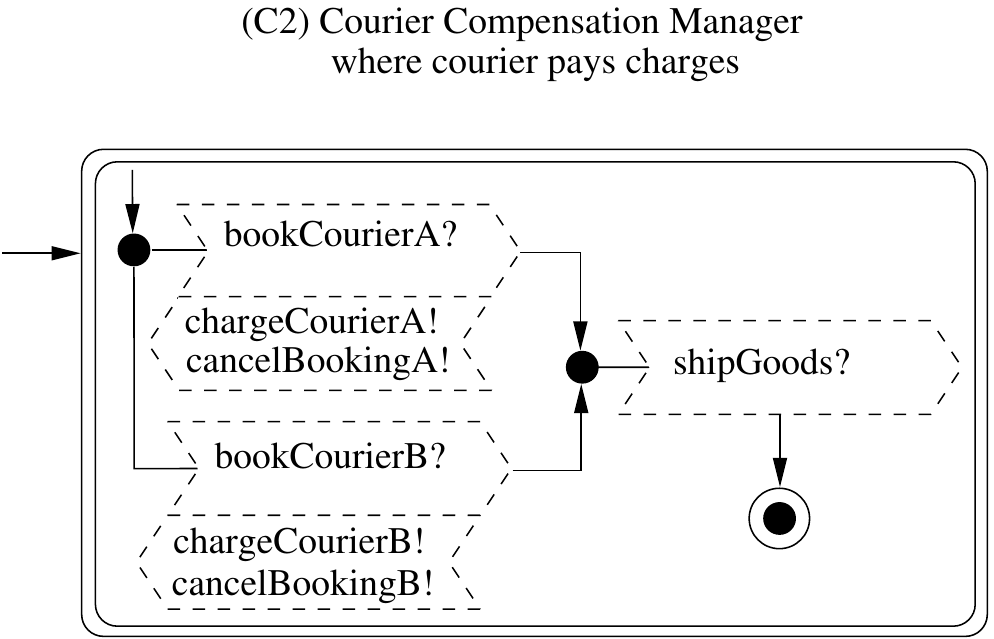}}}\\
\medskip\medskip\medskip
\centering{\scalebox{0.8}{\includegraphics{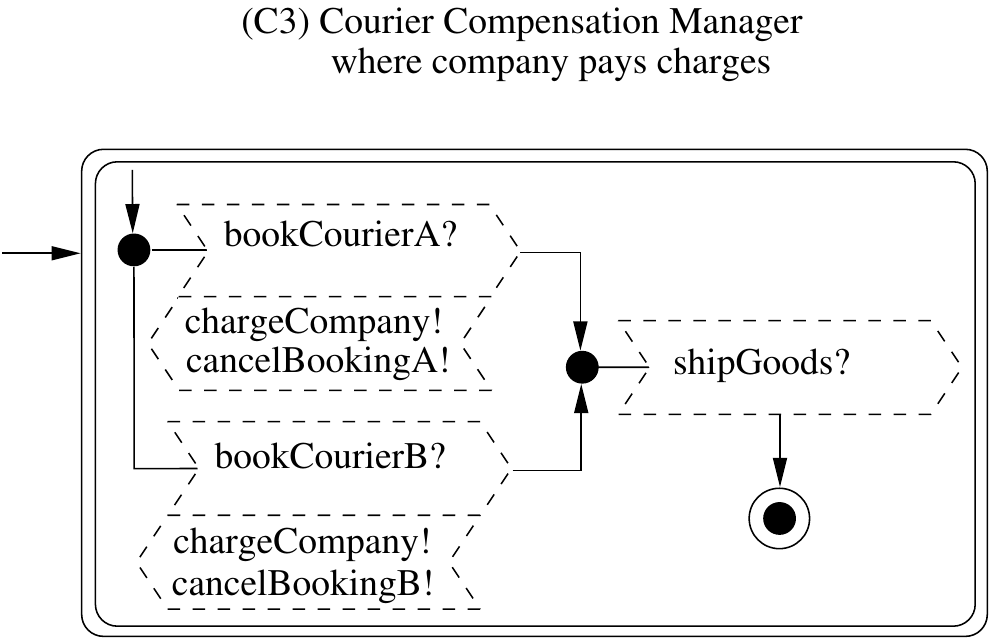}}}\somespace\somespace\somespace\somespace
%
%
\centering{\scalebox{0.8}{\includegraphics{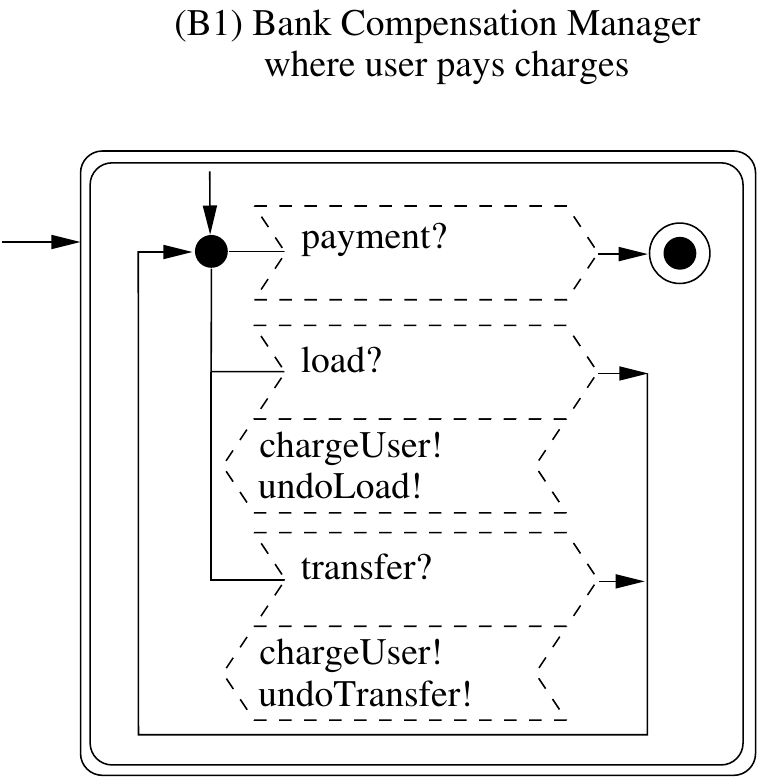}}}\\
\caption[]{Managing compensations for an e-procurement system (continued on the next page)}
\end{figure}

\begin{figure}[t]
\ContinuedFloat
\centering{\scalebox{0.8}{\includegraphics{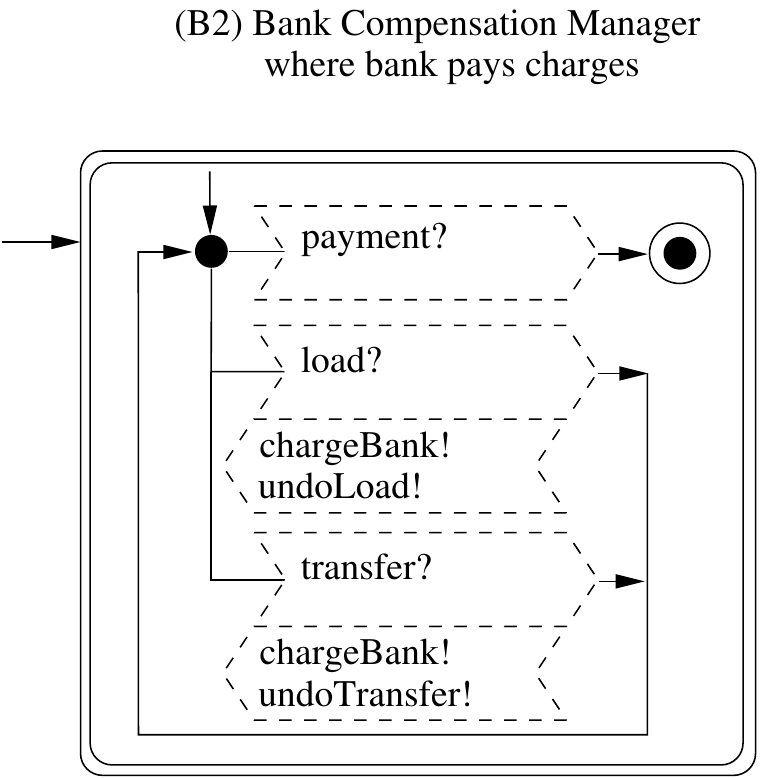}}}\somespace
\centering{\scalebox{0.8}{\includegraphics{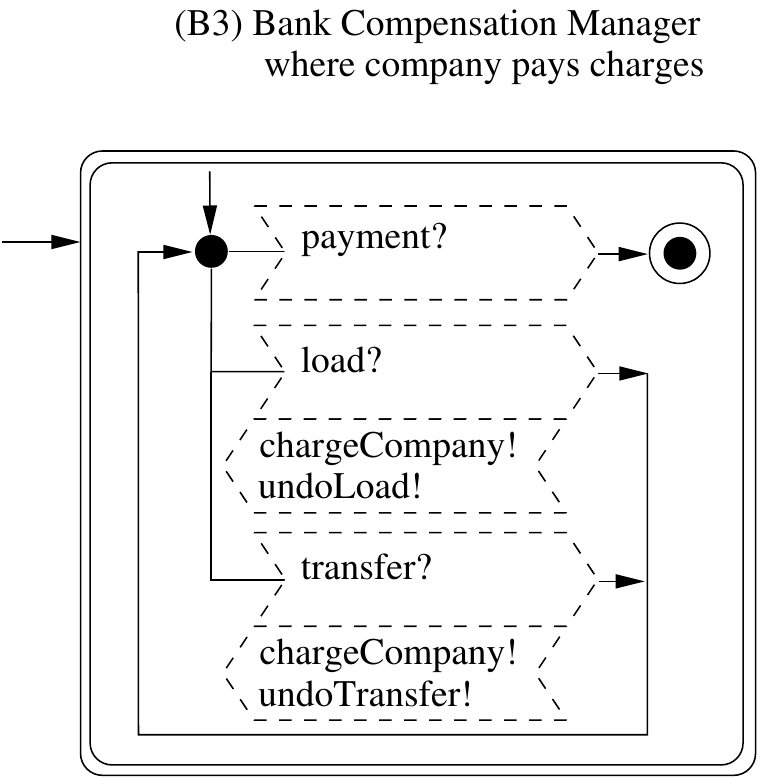}}}\\
\medskip\medskip\medskip
\centering{\scalebox{0.8}{\includegraphics{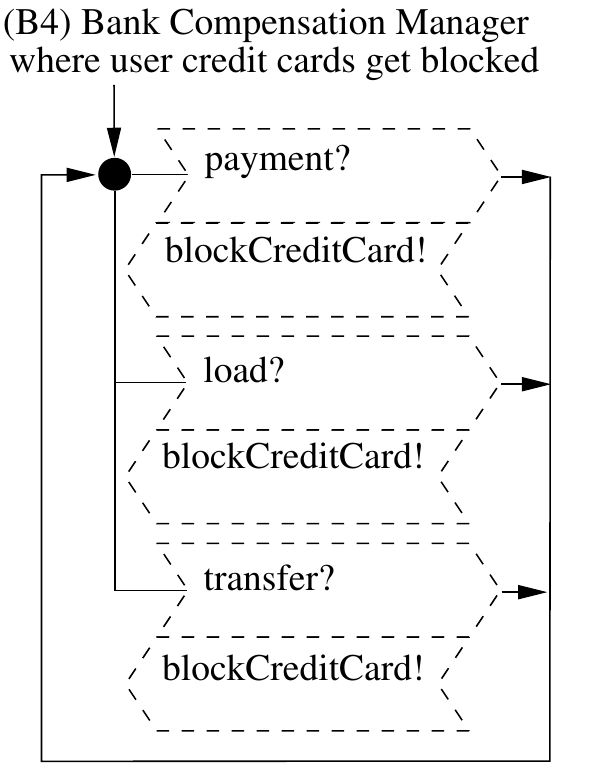}}}\\
\caption[]{Managing compensations for an e-procurement system}
\label{fig:cs-cas}
\end{figure}

\end{landscape}

{\bf Capturing events from system execution}
As a first step, the compensation manager has to be aware of the system's behaviour so that it can react accordingly. In the context of Java systems, the system's behaviour is captured through aspect-oriented programming; with the behaviour typically consisting of method calls, method returns, and exception throws. Different events may be captured depending on the type of system. 

{\bf Passing of relevant system state to the compensation module}
Since the compensation would generally execute on a different (future) system state than the action it is meant to compensate for, it is crucial that any relevant state is passed on for future reference. For example in the case of a payment this might include the account balance at the time of payment (e.g., if a different refund fee applies depending on the balance), and the amount paid. In practical terms, such state would be stored as part of the monitor state and passed on as parameters when invoking the compensation code.

{\bf Enabling the automatic invocation of compensation code}
While the architecture depiction in Figure~\ref{fig:mocp} simply shows a channel output to represent compensation code invocation, an implementation would require a means by which the compensation manager can take over the system execution. To this extent, our standalone monitoring tool {\sc Larva} \cite{CPS09larva} utilises aspect-oriented programming to inject custom code in a Java system. The same approach can be applied in the implementation of the proposed architecture. 


\section{Related Work}\label{sec:rw}

Monitor-oriented programming (MOP) \cite{MJG+11mop} has been proposed as a programming paradigm advocating separation of concerns through monitoring. Somewhat similar to aspect-oriented programming in principle, it differs in that matching occurs through satisfaction of some formal logic rather than source code pattern matching. Inspired by MOP, we have proposed monitor-oriented compensation programming (MOCP) where monitors are used for a specific kind of programming, \ie compensation programming. 
Recall that compensation programming can be split in the programming of four elements: 
$(i)$~\emph{when} --- when to start compensating; 
$(ii)$~\emph{what} --- what system actions to compensate for; 
$(iii)$~\emph{how} --- how to compensate for the designated system actions; and 
$(iv)$~\emph{which} --- which compensation strategy is chosen if there are more than one way of compensating an action or a sequence thereof.  
Thus, a fundamental difference between MOP and MOCP is that while MOP matches a pattern to decide \emph{when} to execute a particular logic, in our case MOCP is concerned not only with the \emph{when} but also with the other questions since compensations are programmed on-the-fly while monitoring and are based on the control flow. MOCP achieves this added expressivity by combining two automata specifications: compensating automata providing the \emph{what} and \emph{how} aspects, and DATEs providing the \emph{when} and \emph{which} elements.

\newcommand{\yes}{\ding{51}}

In the areas of autonomous adaptation and self-healing, one finds much related work  \cite{ACM+07paws,BG11selfsuper,CM07ao4bpel,CDM06scene,EMT07masc,GKM+11,MRD10,SDN07eng,NST+07diamond} 
which has been done in the context of the service-oriented architectures. 
Of these approaches, only a subset \cite{ACM+07paws,BG11selfsuper,CM07ao4bpel,SDN07eng,NST+07diamond} 
support the possibility of executing compensations. 
These works (with the exception of AO4BPEL \cite{CM07ao4bpel}) provide a policy language which is able to separate exception handling (and compensation) concerns from the normal business logic but the policy language itself does not go into the details of compensation programming. In other words, the policy language enables a user to specify \emph{what} and \emph{when} to compensate but it does not provide explicit support for specifying \emph{how} and \emph{which}. Instead, they assume there is a single default compensation for every action. These approaches are referred to as `policy languages' in Table~\ref{tab:comp-exp}. 
On the other hand, AO4BPEL provides an aspect-oriented framework which enables the programming of crosscutting concerns such as compensation programming. However, the language does not provide explicit support for programming compensations and is thus left out of Table~\ref{tab:comp-exp}.\looseness=-1

Most theoretical frameworks which support the specification of compensations \cite{bpmn,
soa/challenges/BMM05,
CGVB+02extending,ES08fct,FR05asm,
GLG+06sock,HZWL08petri,
LPT08blite} 
(referred to as `theoretical frameworks' in Table~\ref{tab:comp-exp}), 
enable the user to specify \emph{what} and \emph{how} to compensate while the compensation is activated automatically upon an external failure/signal. 
On the contrary, other approaches including BPEL and cCSP \cite{bpel2,DBLP:conf/birthday/ButlerHF04} 
 go further to also enable the user to programmatically invoke the compensation activation, \ie deciding \emph{when} to execute the installed compensations. For example in the case of BPEL, compensation is optionally invoked from the exception handler enabling full user control of whether or not to execute compensations. 
 However, all these approaches ignore the \emph{which} question because they assume that there is a single possible compensation strategy.

StAC$_i$ \cite{StAC04BF}, on the other hand, gives full control to the user and several compensation strategies can be programmed concurrently (each with an individual compensation stack) and subsequently the user is allowed to program \emph{when} to run \emph{which} compensation stack. This means that StAC$_i$ supports all four aspects of compensation programming which are supported by MOCP. However, MOCP differs from StAC$_i$ in two fundamental ways:

\begin{itemize}
\item
MOCP clearly separates compensation specification from compensation activation concerns: with compensating automata able to specify \emph{what} and \emph{how} while DATEs can be used for specifying \emph{when} and \emph{which}. We believe that this separation, together with the higher level of abstraction provided through both automaton-based notations, makes compensation programming more manageable. 
\item
Furthermore, MOCP is a monitor-based approach and thus separates compensation concerns from the other concerns. On the other hand, StAC$_i$ requires the programmer to not only program all four aspects of compensation programming but also program the rest of the system concerns including exception handling.
\end{itemize} 
Table~\ref{tab:comp-exp} summarises the capabilities of the reviewed compensation programming approaches.

\begin{table}
\begin{center}
\small
\begin{tabular}{|p{4.5cm}|c|cccc|}
\cline{2-6}
\multicolumn{1}{c|}{}&\multicolumn{1}{c|}{\multirow{2}{*}{System}} & \multicolumn{4}{c|}{Compensation} \\
 \multicolumn{1}{c|}{}&  & when  & what & how & which\\

\hline
Compensating Automata     &          &             & \yes    &   \yes     &        \\
\hline
DATEs                                &         &     \yes   &            &           & \yes \\
\hline
MOCP                                &         &    \yes    &   \yes   &    \yes   & \yes \\

\hline

Policy Languages                &        &   \yes   &  \yes    &            &           \\
\hline
Theoretical Frameworks      & \yes  &          &   \yes    &   \yes       &        \\

\hline
BPEL, cCSP                        & \yes  &    \yes    &   \yes   &    \yes   &        \\

\hline
StAC$_i$                            &\yes   &    \yes    &   \yes   &    \yes   & \yes \\

\hline
\end{tabular}
\end{center}
\caption{Compensation specification approaches vs. expressivity}
\label{tab:comp-exp}
\end{table}

\normalsize

\section{Conclusions}

Whilst compensation programming is frequently understood to answer the questions of what and how to compensate, in more complex scenarios, it is also not straightforward to decide when to trigger compensation strategies, and if a system has more than one compensation strategy, which one to choose. Thus the problem of programming complex compensations can be split into two main aspects: the \emph{what and how} of compensations --- dealing directly with programming what actions to use as compensations --- and the \emph{when and which} of compensations which manages the interplay of normal execution and compensation execution. 

We are currently implementing compensation automata as part of the runtime verification suite Larva, over which we will be implementing our approach and applying it to the case study discussed in this paper where a number of different compensation strategies would need to be maintained concurrently and executed under different contexts. A possible optimisation is to discard compensation strategies which will surely not be used. Taking the example of the case study, as soon as a user is classified as whitelisted, compensation strategies \emph{B2}, \emph{B4}, and \emph{C2} can be discarded, saving the memory and the time to maintain them.

\bibliographystyle{eptcs}
\bibliography{refs}

\begin{thebibliography}{10}
\providecommand{\bibitemdeclare}[2]{}
\providecommand{\surnamestart}{}
\providecommand{\surnameend}{}
\providecommand{\urlprefix}{Available at }
\providecommand{\url}[1]{\texttt{#1}}
\providecommand{\href}[2]{\texttt{#2}}
\providecommand{\urlalt}[2]{\href{#1}{#2}}
\providecommand{\doi}[1]{doi:\urlalt{http://dx.doi.org/#1}{#1}}
\providecommand{\bibinfo}[2]{#2}

\bibitemdeclare{misc}{bpmn}
\bibitem{bpmn}
 (\bibinfo{year}{2008}): \emph{\bibinfo{title}{Business Process Modeling
  Notation, v1.1}}.
\newblock
  \bibinfo{note}{\url{http://www.bpmn.org/Documents/BPMN\_1-1\_Specification.pdf}
  (Last accessed: 2010-02-17)}.

\bibitemdeclare{article}{ACM+07paws}
\bibitem{ACM+07paws}
\bibinfo{author}{Danilo \surnamestart Ardagna\surnameend},
  \bibinfo{author}{Marco \surnamestart Comuzzi\surnameend},
  \bibinfo{author}{Enrico \surnamestart Mussi\surnameend},
  \bibinfo{author}{Barbara \surnamestart Pernici\surnameend} \&
  \bibinfo{author}{Pierluigi \surnamestart Plebani\surnameend}
  (\bibinfo{year}{2007}): \emph{\bibinfo{title}{PAWS: A Framework for Executing
  Adaptive Web-Service Processes}}.
\newblock {\sl \bibinfo{journal}{IEEE Software}} \bibinfo{volume}{24}, pp.
  \bibinfo{pages}{39--46}, \doi{10.1109/MS.2007.174}.

\bibitemdeclare{misc}{bpel2}
\bibitem{bpel2}
\bibinfo{author}{A.~\surnamestart Arkin\surnameend},
  \bibinfo{author}{S.~\surnamestart Askary\surnameend},
  \bibinfo{author}{B.~\surnamestart Bloch\surnameend},
  \bibinfo{author}{F.~\surnamestart Curbera\surnameend},
  \bibinfo{author}{Y.~\surnamestart Goland\surnameend},
  \bibinfo{author}{N.~\surnamestart Kartha\surnameend}, \bibinfo{author}{C.~K.
  \surnamestart Liu\surnameend}, \bibinfo{author}{S.~\surnamestart
  Thatte\surnameend}, \bibinfo{author}{P.~\surnamestart Yendluri\surnameend} \&
  \bibinfo{author}{A.~\surnamestart Yiu\surnameend} (\bibinfo{year}{2007}):
  \emph{\bibinfo{title}{Web Services Business Process Execution Language
  Version 2.0}}.
\newblock \bibinfo{note}{OASIS Standard. Available at:
  http://docs.oasis-open.org/wsbpel/2.0/wsbpel-v2.0.pdf (Last accessed:
  2010-02-17)}.

\bibitemdeclare{article}{BG11selfsuper}
\bibitem{BG11selfsuper}
\bibinfo{author}{L.~\surnamestart Baresi\surnameend} \&
  \bibinfo{author}{S.~\surnamestart Guinea\surnameend} (\bibinfo{year}{2011}):
  \emph{\bibinfo{title}{Self-Supervising BPEL Processes}}.
\newblock {\sl \bibinfo{journal}{Software Engineering, IEEE Transactions on}}
  \bibinfo{volume}{37}(\bibinfo{number}{2}), pp. \bibinfo{pages}{247 --263},
  \doi{10.1109/TSE.2010.37}.

\bibitemdeclare{inproceedings}{soa/challenges/BMM05}
\bibitem{soa/challenges/BMM05}
\bibinfo{author}{Roberto \surnamestart Bruni\surnameend},
  \bibinfo{author}{Hern\'{a}n \surnamestart Melgratti\surnameend} \&
  \bibinfo{author}{Ugo \surnamestart Montanari\surnameend}
  (\bibinfo{year}{2005}): \emph{\bibinfo{title}{Theoretical foundations for
  compensations in flow composition languages}}.
\newblock In: {\sl \bibinfo{booktitle}{POPL}}, \bibinfo{publisher}{ACM}, pp.
  \bibinfo{pages}{209--220}, \doi{10.1145/1040305.1040323}.

\bibitemdeclare{inproceedings}{StAC04BF}
\bibitem{StAC04BF}
\bibinfo{author}{Michael~J. \surnamestart Butler\surnameend} \&
  \bibinfo{author}{Carla \surnamestart Ferreira\surnameend}
  (\bibinfo{year}{2004}): \emph{\bibinfo{title}{An Operational Semantics for
  StAC, a Language for Modelling Long-Running Business Transactions}}.
\newblock In: {\sl \bibinfo{booktitle}{COORDINATION}}, {\sl
  \bibinfo{series}{LNCS}} \bibinfo{volume}{2949},
  \bibinfo{publisher}{Springer}, pp. \bibinfo{pages}{87--104},
  \doi{10.1007/978-3-540-24634-3\_9}.

\bibitemdeclare{inproceedings}{DBLP:conf/birthday/ButlerHF04}
\bibitem{DBLP:conf/birthday/ButlerHF04}
\bibinfo{author}{Michael~J. \surnamestart Butler\surnameend},
  \bibinfo{author}{C.~A.~R. \surnamestart Hoare\surnameend} \&
  \bibinfo{author}{Carla \surnamestart Ferreira\surnameend}
  (\bibinfo{year}{2004}): \emph{\bibinfo{title}{A Trace Semantics for
  Long-Running Transactions}}.
\newblock In: {\sl \bibinfo{booktitle}{25 Years Communicating Sequential
  Processes}}, \bibinfo{series}{LNCS}, \bibinfo{publisher}{Springer}, pp.
  \bibinfo{pages}{133--150}, \doi{10.1007/11423348\_8}.

\bibitemdeclare{article}{CM07ao4bpel}
\bibitem{CM07ao4bpel}
\bibinfo{author}{Anis \surnamestart Charfi\surnameend} \& \bibinfo{author}{Mira
  \surnamestart Mezini\surnameend} (\bibinfo{year}{2007}):
  \emph{\bibinfo{title}{AO4BPEL: An Aspect-oriented Extension to BPEL}}.
\newblock {\sl \bibinfo{journal}{World Wide Web}}
  \bibinfo{volume}{10}(\bibinfo{number}{3}), pp. \bibinfo{pages}{309--344},
  \doi{10.1007/s11280-006-0016-3}.

\bibitemdeclare{article}{CGVB+02extending}
\bibitem{CGVB+02extending}
\bibinfo{author}{M.~\surnamestart Chessell\surnameend},
  \bibinfo{author}{C.~\surnamestart Griffin\surnameend},
  \bibinfo{author}{D.~\surnamestart Vines\surnameend},
  \bibinfo{author}{M.~\surnamestart Butler\surnameend},
  \bibinfo{author}{C.~\surnamestart Ferreira\surnameend} \&
  \bibinfo{author}{P.~\surnamestart Henderson\surnameend}
  (\bibinfo{year}{2002}): \emph{\bibinfo{title}{Extending the concept of
  transaction compensation}}.
\newblock {\sl \bibinfo{journal}{IBM Systems Journal}}
  \bibinfo{volume}{41}(\bibinfo{number}{4}), pp. \bibinfo{pages}{743--758},
  \doi{10.1147/sj.414.0743}.

\bibitemdeclare{article}{cp13cas}
\bibitem{cp13cas}
\bibinfo{author}{Christian \surnamestart Colombo\surnameend} \&
  \bibinfo{author}{Gordon \surnamestart Pace\surnameend}
  (\bibinfo{year}{2013}): \emph{\bibinfo{title}{Monitor-Oriented Compensation
  Programming Through Compensating Automata}}.
\newblock {\sl \bibinfo{journal}{ECEASST}} \bibinfo{volume}{58}.

\bibitemdeclare{article}{survey}
\bibitem{survey}
\bibinfo{author}{Christian \surnamestart Colombo\surnameend} \&
  \bibinfo{author}{Gordon \surnamestart Pace\surnameend}
  (\bibinfo{year}{2013}): \emph{\bibinfo{title}{Recovery within Long Running
  Transactions}}.
\newblock {\sl \bibinfo{journal}{ACM Computing Surveys}} \bibinfo{volume}{45},
  \doi{10.1145/2480741.2480745}.

\bibitemdeclare{article}{CPA12async}
\bibitem{CPA12async}
\bibinfo{author}{Christian \surnamestart Colombo\surnameend},
  \bibinfo{author}{Gordon \surnamestart Pace\surnameend} \&
  \bibinfo{author}{Patrick \surnamestart Abela\surnameend}
  (\bibinfo{year}{2012}): \emph{\bibinfo{title}{Safer asynchronous runtime
  monitoring using compensations}}.
\newblock {\sl \bibinfo{journal}{Formal Methods in System Design}}
  \bibinfo{volume}{41}(\bibinfo{number}{3}), pp. \bibinfo{pages}{269--294},
  \doi{10.1007/s10703-012-0142-8}.

\bibitemdeclare{inproceedings}{CGG08FMICS}
\bibitem{CGG08FMICS}
\bibinfo{author}{Christian \surnamestart Colombo\surnameend},
  \bibinfo{author}{Gordon~J. \surnamestart Pace\surnameend} \&
  \bibinfo{author}{Gerardo \surnamestart Schneider\surnameend}
  (\bibinfo{year}{2008}): \emph{\bibinfo{title}{Dynamic Event-Based Runtime
  Monitoring of Real-Time and Contextual Properties}}.
\newblock In: {\sl \bibinfo{booktitle}{FMICS}}, {\sl \bibinfo{series}{LNCS}}
  \bibinfo{volume}{5596}, pp. \bibinfo{pages}{135--149},
  \doi{10.1007/978-3-642-03240-0\_13}.

\bibitemdeclare{inproceedings}{CPS09larva}
\bibitem{CPS09larva}
\bibinfo{author}{Christian \surnamestart Colombo\surnameend},
  \bibinfo{author}{Gordon~J. \surnamestart Pace\surnameend} \&
  \bibinfo{author}{Gerardo \surnamestart Schneider\surnameend}
  (\bibinfo{year}{2009}): \emph{\bibinfo{title}{LARVA --- Safer Monitoring of
  Real-Time Java Programs (Tool Paper)}}.
\newblock In: {\sl \bibinfo{booktitle}{SEFM}}, \bibinfo{publisher}{IEEE}, pp.
  \bibinfo{pages}{33--37}, \doi{10.1109/SEFM.2009.13}.

\bibitemdeclare{inproceedings}{CDM06scene}
\bibitem{CDM06scene}
\bibinfo{author}{Massimiliano \surnamestart Colombo\surnameend},
  \bibinfo{author}{Elisabetta \surnamestart Di~Nitto\surnameend} \&
  \bibinfo{author}{Marco \surnamestart Mauri\surnameend}
  (\bibinfo{year}{2006}): \emph{\bibinfo{title}{SCENE: a service composition
  execution environment supporting dynamic changes disciplined through rules}}.
\newblock In: {\sl \bibinfo{booktitle}{ICSOC}}, \bibinfo{publisher}{Springer},
  pp. \bibinfo{pages}{191--202}, \doi{10.1007/11948148\_16}.

\bibitemdeclare{inproceedings}{DAVI72}
\bibitem{DAVI72}
\bibinfo{author}{Charles~T. \surnamestart Davies\surnameend, Jr.}
  (\bibinfo{year}{1973}): \emph{\bibinfo{title}{Recovery semantics for a
  {DB/DC} system}}.
\newblock In: {\sl \bibinfo{booktitle}{ACM annual conference}},
  \bibinfo{publisher}{ACM}, pp. \bibinfo{pages}{136--141},
  \doi{10.1145/800192.805694}.

\bibitemdeclare{inproceedings}{ES08fct}
\bibitem{ES08fct}
\bibinfo{author}{Christian \surnamestart Eisentraut\surnameend} \&
  \bibinfo{author}{David \surnamestart Spieler\surnameend}
  (\bibinfo{year}{2008}): \emph{\bibinfo{title}{Fault, Compensation and
  Termination in {WS-BPEL} 2.0 - A Comparative Analysis}}.
\newblock In: {\sl \bibinfo{booktitle}{WS-FM}}, {\sl \bibinfo{series}{LNCS}}
  \bibinfo{volume}{5387}, \bibinfo{publisher}{Springer}, pp.
  \bibinfo{pages}{107--126}, \doi{10.1007/978-3-642-01364-5\_7}.

\bibitemdeclare{inproceedings}{EMT07masc}
\bibitem{EMT07masc}
\bibinfo{author}{Abdelkarim \surnamestart Erradi\surnameend},
  \bibinfo{author}{Piyush \surnamestart Maheshwari\surnameend} \&
  \bibinfo{author}{Vladimir \surnamestart Tosic\surnameend}
  (\bibinfo{year}{2007}): \emph{\bibinfo{title}{WS-Policy based Monitoring of
  Composite Web Services}}.
\newblock In: {\sl \bibinfo{booktitle}{ECOWS}}, \bibinfo{publisher}{IEEE}, pp.
  \bibinfo{pages}{99--108}, \doi{10.1109/ECOWS.2007.31}.

\bibitemdeclare{inproceedings}{FR05asm}
\bibitem{FR05asm}
\bibinfo{author}{Dirk \surnamestart Fahland\surnameend} \&
  \bibinfo{author}{Wolfgang \surnamestart Reisig\surnameend}
  (\bibinfo{year}{2005}): \emph{\bibinfo{title}{{ASM}-based Semantics for
  {BPEL}: The Negative Control Flow}}.
\newblock In: {\sl \bibinfo{booktitle}{ASM}}, pp. \bibinfo{pages}{131--152}.

\bibitemdeclare{inproceedings}{GFJK03notenough}
\bibitem{GFJK03notenough}
\bibinfo{author}{Paul \surnamestart Greenfield\surnameend},
  \bibinfo{author}{Alan \surnamestart Fekete\surnameend},
  \bibinfo{author}{Julian \surnamestart Jang\surnameend} \&
  \bibinfo{author}{Dean \surnamestart Kuo\surnameend} (\bibinfo{year}{2003}):
  \emph{\bibinfo{title}{Compensation is Not Enough}}.
\newblock In: {\sl \bibinfo{booktitle}{EDOC}}, \bibinfo{publisher}{IEEE}, pp.
  \bibinfo{pages}{232--239}, \doi{10.1109/EDOC.2003.1233852}.

\bibitemdeclare{inproceedings}{GLG+06sock}
\bibitem{GLG+06sock}
\bibinfo{author}{Claudio \surnamestart Guidi\surnameend},
  \bibinfo{author}{Roberto \surnamestart Lucchi\surnameend},
  \bibinfo{author}{Roberto \surnamestart Gorrieri\surnameend},
  \bibinfo{author}{Nadia \surnamestart Busi\surnameend} \&
  \bibinfo{author}{Gianluigi \surnamestart Zavattaro\surnameend}
  (\bibinfo{year}{2006}): \emph{\bibinfo{title}{{SOCK}: A Calculus for Service
  Oriented Computing}}.
\newblock In: {\sl \bibinfo{booktitle}{ICSOC}}, {\sl \bibinfo{series}{LNCS}}
  \bibinfo{volume}{4294}, \bibinfo{publisher}{Springer}, pp.
  \bibinfo{pages}{327--338}, \doi{10.1007/11948148\_27}.

\bibitemdeclare{incollection}{GKM+11}
\bibitem{GKM+11}
\bibinfo{author}{Sam \surnamestart Guinea\surnameend}, \bibinfo{author}{Gabor
  \surnamestart Kecskemeti\surnameend}, \bibinfo{author}{Annapaola
  \surnamestart Marconi\surnameend} \& \bibinfo{author}{Branimir \surnamestart
  Wetzstein\surnameend} (\bibinfo{year}{2011}):
  \emph{\bibinfo{title}{Multi-layered Monitoring and Adaptation}}.
\newblock In: {\sl \bibinfo{booktitle}{Service-Oriented Computing}}, {\sl
  \bibinfo{series}{LNCS}} \bibinfo{volume}{7084},
  \bibinfo{publisher}{Springer}, pp. \bibinfo{pages}{359--373},
  \doi{10.1007/978-3-642-25535-9\_24}.

\bibitemdeclare{inproceedings}{HZWL08petri}
\bibitem{HZWL08petri}
\bibinfo{author}{Yanxiang \surnamestart He\surnameend}, \bibinfo{author}{Liang
  \surnamestart Zhao\surnameend}, \bibinfo{author}{Zhao \surnamestart
  Wu\surnameend} \& \bibinfo{author}{Fei \surnamestart Li\surnameend}
  (\bibinfo{year}{2008}): \emph{\bibinfo{title}{Formal Modeling of Transaction
  Behavior in {WS-BPEL}}}.
\newblock In: {\sl \bibinfo{booktitle}{CSSE}}, \bibinfo{publisher}{IEEE}, pp.
  \bibinfo{pages}{490--494}, \doi{10.1109/CSSE.2008.873}.

\bibitemdeclare{inproceedings}{LPT08blite}
\bibitem{LPT08blite}
\bibinfo{author}{Alessandro \surnamestart Lapadula\surnameend},
  \bibinfo{author}{Rosario \surnamestart Pugliese\surnameend} \&
  \bibinfo{author}{Francesco \surnamestart Tiezzi\surnameend}
  (\bibinfo{year}{2008}): \emph{\bibinfo{title}{A Formal Account of
  {WS-BPEL}}}.
\newblock In: {\sl \bibinfo{booktitle}{COORDINATION}}, {\sl
  \bibinfo{series}{LNCS}} \bibinfo{volume}{5052},
  \bibinfo{publisher}{Springer}, pp. \bibinfo{pages}{199--215},
  \doi{10.1007/978-3-540-68265-3\_13}.

\bibitemdeclare{article}{MJG+11mop}
\bibitem{MJG+11mop}
\bibinfo{author}{Patrick~O'Neil \surnamestart Meredith\surnameend},
  \bibinfo{author}{Dongyun \surnamestart Jin\surnameend},
  \bibinfo{author}{Dennis \surnamestart Griffith\surnameend},
  \bibinfo{author}{Feng \surnamestart Chen\surnameend} \&
  \bibinfo{author}{Grigore \surnamestart Ro\c{s}u\surnameend}
  (\bibinfo{year}{2012}): \emph{\bibinfo{title}{An Overview of the {MOP}
  Runtime Verification Framework}}.
\newblock {\sl \bibinfo{journal}{Journal on Software Tools for Technology
  Transfer}} \bibinfo{volume}{14}(\bibinfo{number}{3}),
  \doi{10.1007/s10009-011-0198-6}.

\bibitemdeclare{inproceedings}{MRD10}
\bibitem{MRD10}
\bibinfo{author}{Oliver \surnamestart Moser\surnameend},
  \bibinfo{author}{Florian \surnamestart Rosenberg\surnameend} \&
  \bibinfo{author}{Schahram \surnamestart Dustdar\surnameend}
  (\bibinfo{year}{2010}): \emph{\bibinfo{title}{Event driven monitoring for
  service composition infrastructures}}.
\newblock In: {\sl \bibinfo{booktitle}{WISE}}, \bibinfo{publisher}{Springer},
  pp. \bibinfo{pages}{38--51}, \doi{10.1007/978-3-642-17616-6\_6}.

\bibitemdeclare{inproceedings}{NFG+05sowl}
\bibitem{NFG+05sowl}
\bibinfo{author}{Surya \surnamestart Nepal\surnameend}, \bibinfo{author}{Alan
  \surnamestart Fekete\surnameend}, \bibinfo{author}{Paul \surnamestart
  Greenfield\surnameend}, \bibinfo{author}{Julian \surnamestart
  Jang\surnameend}, \bibinfo{author}{Dean \surnamestart Kuo\surnameend} \&
  \bibinfo{author}{Tony \surnamestart Shi\surnameend} (\bibinfo{year}{2005}):
  \emph{\bibinfo{title}{A service-oriented workflow language for robust
  interacting applications}}.
\newblock In: {\sl \bibinfo{booktitle}{On the Move to Meaningful Internet
  Systems - Part I}}, \bibinfo{publisher}{Springer}, pp.
  \bibinfo{pages}{40--58}, \doi{10.1007/11575771\_6}.

\bibitemdeclare{article}{RICSD78}
\bibitem{RICSD78}
\bibinfo{author}{B.~\surnamestart Randell\surnameend},
  \bibinfo{author}{P.~\surnamestart Lee\surnameend} \& \bibinfo{author}{P.~C.
  \surnamestart Treleaven\surnameend} (\bibinfo{year}{1978}):
  \emph{\bibinfo{title}{Reliability Issues in Computing System Design}}.
\newblock {\sl \bibinfo{journal}{ACM Computing Surveys}} \bibinfo{volume}{10},
  pp. \bibinfo{pages}{123--165}, \doi{10.1145/356725.356729}.

\bibitemdeclare{inproceedings}{SDN07eng}
\bibitem{SDN07eng}
\bibinfo{author}{Michael \surnamestart Sch\"{a}fer\surnameend},
  \bibinfo{author}{Peter \surnamestart Dolog\surnameend} \&
  \bibinfo{author}{Wolfgang \surnamestart Nejdl\surnameend}
  (\bibinfo{year}{2007}): \emph{\bibinfo{title}{Engineering compensations in
  web service environment}}.
\newblock In: {\sl \bibinfo{booktitle}{ICWE}}, \bibinfo{publisher}{Springer},
  pp. \bibinfo{pages}{32--46}.

\bibitemdeclare{incollection}{NST+07diamond}
\bibitem{NST+07diamond}
\bibinfo{author}{Nick~Amirreza \surnamestart Tahamtan\surnameend} \&
  \bibinfo{author}{WS-Diamond \surnamestart team\surnameend}
  (\bibinfo{year}{2007}): \emph{\bibinfo{title}{WS-DIAMOND: Web Services -
  DIAgnosability, MONitoring and Diagnosis}}.
\newblock In: {\sl \bibinfo{booktitle}{E. di Nitto, A. Sassen, P.Traverso and
  A. Zwegers (Eds), At your service, Chapter 9}}, \bibinfo{publisher}{MIT
  Press}.

\end{thebibliography}

\end{document}